\documentclass{article}

\usepackage{amssymb,amsfonts,amsmath}
\usepackage{cite,enumerate,float}
\usepackage{color}
\usepackage{tikz}
\usetikzlibrary{arrows,snakes,backgrounds}

\def\be{\begin{eqnarray}}
\def\ee{\end{eqnarray}}
\def\nn{\nonumber}

\definecolor{red}{rgb}{1,0,0}
\definecolor{orange}{rgb}{1,0.5,0}
\definecolor{violet}{rgb}{0.7,0,1}

\hoffset= - 1.0in         

\begin{document}

\title{\vspace{1.5cm}\bf
Twisted Baker-Akhiezer function from determinants
}

\author{
A. Mironov$^{b,c,d,}$\footnote{mironov@lpi.ru,mironov@itep.ru},
A. Morozov$^{a,c,d,}$\footnote{morozov@itep.ru},
A. Popolitov$^{a,c,d,}$\footnote{popolit@gmail.com}
}

\date{ }

\maketitle

\vspace{-6cm}

\begin{center}
  \hfill MIPT/TH-08/25\\
  \hfill FIAN/TD-06/25\\
  \hfill ITEP/TH-10/25\\
  \hfill IITP/TH-08/25
\end{center}

\vspace{4.5cm}

\begin{center}
$^a$ {\small {\it MIPT, Dolgoprudny, 141701, Russia}}\\
$^b$ {\small {\it Lebedev Physics Institute, Moscow 119991, Russia}}\\
$^c$ {\small {\it NRC ``Kurchatov Institute", 123182, Moscow, Russia}}\\
$^d$ {\small {\it Institute for Information Transmission Problems, Moscow 127994, Russia}}
\end{center}

\vspace{.1cm}

\begin{abstract}
General description of eigenfunctions of integrable Hamiltonians associated with the integer rays of Ding-
Iohara-Miki (DIM) algebra, is provided by the theory of Chalykh Baker-Akhiezer functions (BAF) defined
as solutions to a simply looking linear system. Solutions themselves are somewhat complicated, but much
simpler than they could. It is because of simultaneous partial factorization of all the determinants, entering
Cramer's rule. This is a conspiracy responsible for a relative simplicity of the Macdonald polynomials and of
the Noumi-Shirashi functions, and it is further continued to all integer DIM rays. Still, factorization is only
partial, moreover, there are different branches and abrupt jumps between them. We explain this feature
of Cramer's rule in an example of a matrix that defines BAF and exhibits a non-analytical dependence on parameters.
Moreover, the matrix is such that there is no natural expansion around non-degenerate approximations,
which causes an unexpected complexity of formulas.
\end{abstract}

\bigskip

\newcommand\smallpar[1]{
  \noindent $\bullet$ \textbf{#1}
}

\section{Introduction}

In \cite{MMP1,MMP2}, following the original suggestion of \cite{Cha,CE}, we investigated
a funny system of $ma$ linear equations
\be
Q^j \cdot \Psi^{(a)}_m \left(\epsilon^{-s} K^j \right) =
\epsilon^{sj} \cdot \Psi^{(a)}_m \left(\epsilon^{-s} K^{-j}\right)
\label{linsys}
\ee
with $\epsilon:=e^{\frac{2\pi i}{a}}$, $s=1,\ldots,a$ and $j=1,\ldots,m$
for the $ma$ coefficients $\psi^{(a)}_{m,k}$ of
\be
\Psi^{(a)}_m(Z) := Z^{\frac{ ma}{2}}\sum_{k=0}^{am} Z^{-k} \psi^{(a)}_{m,k}
\label{Psi}
\ee
It turns out that, at $a=1$, this system describes the rank one Baker-Akhiezer functions (BAF), which are eigenfunctions of the two-particle trigonometric Ruijsenaars Hamiltonians \cite{Mac,Rui} (at particular values of $Q$, they have Macdonald polynomials as eigenfunctions \cite{Mac}),
and, at other natural $a$, it does the same \cite{CF,MMP1} for the $a$-th integer ray \cite{MMP} associated with
a commutative subalgebra in DIM \cite{DIM}, thus
providing a non-trivial generalization of Macdonald theory \cite{Mac,Cher}.

This is an unusually simple look at Macdonald theory, and it is not a surprise that it provides
these far-going generalizations.
It is also a non-trivial look, because relation to the Macdonald polynomials is somewhat peculiar:
  ${\cal M}_{[r]}[x_1,x_2]$  with $r>2m$ reduces at $t=q^{-m}$  to a sum
$x_1^r \Psi^{(1)}_{m}(x_2/x_1) + x_2^r \Psi^{(1)}_{m}(x_1/x_2)$
with $Q=q^{r-m}$ and $K=q$.
Such an approach deals with the original Macdonald polynomials at arbitrary $t$ as with an analytical continuation
of $\Psi^{(1)}_{m}$, but instead the latter ones satisfy a nearly trivial system (\ref{linsys}),
while the trigonometric Ruijsenaars Hamiltonians for the Macdonald polynomials {\it per se} are pretty complicated.
Nothing to say about generalization to $a>1$, where the Hamiltonians are even not that simple to construct
in an explicit form \cite{MMP2}.
At $a>1$, the association of parameters is $\boxed{ Q=q^{\lambda/a}, \ K=q^{1/a},}$, but since we do not need the standard $q$ and $\lambda$,
we mostly use (for exception of a couple of formulas in sec.3.3) throughout the text just $Q$ and $K$ to simplify formulas.

The system (\ref{linsys}) has a simple generalization to an arbitrary $N$ counting the number of
variables $x_1,\ldots,x_N$ in BAF. The associated Macdonald polynomials are symmetric polynomials of these variables, and extending their set from $N=2$ to arbitrary $N$ allows one to deal with the Macdonald polynomials and their $a$-generalizations (which is called $a$-twist) in non-symmetric representations. The BAF at $a\ne 1$ are called $a$-twist BAF \cite{CE}.
The consideration below can be easily extended to an arbitrary $N$, but it has some peculiarities
outlined in sec.3 of \cite{MMP1}, hence, we do not consider $N>2$ in what follows to avoid additional complications.

Instead we discuss a straightforward solution to (\ref{linsys}) with the help of Cramer's rule,
which should extend the limited set of explicit examples of \cite{Cha,MMP1,MMP2} and sheds more light on their
structure, which there was left somewhat obscure.
It also provides additional colors for the seemingly mysterious theory of Noumi-Shiraishi functions \cite{NS},
which were shown \cite{MMP3} to have a direct relation to the same problem (\ref{linsys}).

\section{Introductory examples}

At ${\bf a=1,m=1}$, the system (\ref{linsys}) reduces to a single equation
\be
Q\sqrt{\frac{K}{\epsilon}}\left(\psi^{(1)}_{1,0}+\frac{\epsilon}{K}\psi^{(1)}_{1,1}\right)
= \frac{\epsilon}{\sqrt{\epsilon K}} \left(\psi^{(1)}_{1,0}+\epsilon K \psi^{(1)}_{1,1}\right)
\ee
with a solution
\be
\frac{\psi^{(1)}_{1,1}}{\psi^{(1)}_{1,0}} = \frac{QK-\epsilon}{\epsilon(\epsilon K-Q)}
\ \stackrel{\epsilon =1}{\Longrightarrow}\ \frac{QK-1}{K-Q}
\ee

\bigskip

At ${\bf a=1,m=2}$, there are already two equations:
\be
{\frac{QK}{\epsilon}}\left(\psi^{(1)}_{2,0}+\frac{\epsilon}{K}\psi^{(1)}_{2,1}+\frac{\epsilon^2}{K^2}\psi^{(1)}_{2,2}\right)
= \frac{  \left(\psi^{(1)}_{2,0}+\epsilon K \psi^{(1)}_{2,1}+\epsilon^2K^2 \psi^{(1)}_{[2,2}\right)}{K}
\nn \\
{\frac{Q^2K^2}{\epsilon}}\left(\psi^{(1)}_{2,0}+\frac{\epsilon}{K^2}\psi^{(1)}_{2,1}+\frac{\epsilon^2}{K^4}\psi^{(1)}_{2,2}\right)
= \frac{\epsilon  \left(\psi^{(1)}_{2,0}+\epsilon K^2 \psi^{(1)}_{2,1}+\epsilon^2K^4 \psi^{(1)}_{[2,2}\right)}{K^2}
\ee
which implies
\be
\frac{\psi^{(1)}_{2,1}}{\psi^{(1)}_{2,0}} = \frac{(K+1)(K^2Q-\epsilon)}{\epsilon K (\epsilon K-Q)}
\ \stackrel{\epsilon =1}{\Longrightarrow} \ \frac{(K+1)(K^2Q-1)}{K(K-Q)}
\nn \\
\frac{\psi^{(1)}_{2,2}}{\psi^{(1)}_{2,0}} = \frac{(QK-\epsilon)(QK^2-\epsilon)}{\epsilon^2(\epsilon K-Q) (\epsilon K^2-Q)}
\ \stackrel{\epsilon =1}{\Longrightarrow}  \ \frac{(QK-1)(QK^2-1)}{(K-Q)(K^2-Q)}
\ee

\bigskip

At ${\bf a=2,m=1}$, the number of equations is again two, but the second equation is slightly different:
\be
{\frac{QK}{\epsilon}}\left(\psi^{(2)}_{1,0}+\frac{\epsilon}{K}\psi^{(2)}_{1,1}+\frac{\epsilon^2}{K^2}\psi^{(2)}_{1,2}\right)
= \frac{  \left(\psi^{(2)}_{1,0}+\epsilon K \psi^{(2)}_{1,1}+\epsilon^2K^2 \psi^{(2)}_{[1,2}\right)}{K}
\nn \\
{\frac{QK}{\epsilon^2}}\left(\psi^{(2)}_{1,0}+\frac{\epsilon^2}{K}\psi^{(2)}_{1,1}+\frac{\epsilon^4}{K^2}\psi^{(2)}_{1,2}\right)
= \frac{  \left(\psi^{(2)}_{1,0}+\epsilon^2 K \psi^{(2)}_{1,1}+\epsilon^4K^2 \psi^{(1)}_{[2,2}\right)}{K}
\ee
However, the solutions are drastically different:
\be
\frac{\psi^{(2)}_{1,1}}{\psi^{(2)}_{1,0}} =
\frac{K^4Q(\epsilon+\epsilon^2+\epsilon^3)-K^2(Q^2+Q^2\epsilon+\epsilon^3+\epsilon^4)+Q\epsilon^2}
{\epsilon^2 K \Big(K^2(\epsilon^3-Q\epsilon^2-Q\epsilon)+Q^2\Big)}
\ \stackrel{\epsilon =-1}{\Longrightarrow} \   \frac{Q(K^4-1)}{K(K^2-Q^2)}
\nn \\
\frac{\psi^{(2)}_{1,2}}{\psi^{(2)}_{1,0}} =
- \frac{K^2(Q\epsilon^2+Q\epsilon-Q^2)-\epsilon^3 }
{\epsilon^3 \Big(K^2(\epsilon^3-Q\epsilon^2-Q\epsilon)+Q^2\Big)}
\ \stackrel{\epsilon =-1}{\Longrightarrow}  \  \frac{Q^2K^2-1}{K^2-Q^2}
\ee
and get factorized only at the proper $\epsilon=-1$.

\bigskip

One has to generalize these three examples and to make calculation equally simple for all $a$ and $m$.

\section{Cramer's rule for (\ref{linsys}) }

One can look at (\ref{linsys}) as at the set of equations
\be
\sum_{k=0}^{am} \epsilon^{sk} \left(\frac{u^j}{\epsilon^{sj}} K^{-jk} - K^{jk}\right)\psi_{m,k}^{(a)} = 0
\ee
with $j=1,\ldots,m$, $s=1,\ldots ,a$ and $\boxed{u=K^{ma}Q = q^{m+\frac{\lambda}{a}}}$, or, equivalently,
\be\label{Eq}
\sum_{k=0}^{ma} M_{l,k} \psi^{(a)}_{m,k} = 0
\ee
with the multi-index $l = (j,s)=1,\ldots,ma$.
We usually order the labels $l$ in the following way: $l = (s-1)m+j$.
In other words, we fix $s$ and then change $j$, then take the next $s$ and again change $j$, and so on, i.e. $j = l \ {\bf mod}(m) $.

Solution to (\ref{Eq}) is provided by Cramer's rule:
 \be
\frac{\psi^{(a)}_{m,k}}{\psi^{(a)}_{m,0}} = (-)^k\cdot \frac{  {\rm det}(M^{(k)})}{{\rm det}( M^{(0)})}
\ee
where $M^{(k)}$ denotes the matrix with the $k$-th column removed.
The point is that the determinants at the r.h.s. are very complicated,
but get nicely factorized exactly at $\epsilon=e^{\frac{2\pi i}{a}}$.
The question is whether they can be evaluated explicitly at arbitrary $a$.

\subsection{$a=1$}

In this case \cite[sec.4.1]{Cha}, the factorization holds at arbitrary $\epsilon$, not obligatory restricted to be $\epsilon=e^{\frac{2\pi i}{a}}$, though for the BAF we need just $\epsilon=1$:

\be
{\rm det} (M^{(k)}) = (-1)^k{\rm det} (M^{(0)})
\cdot
\boxed{\epsilon^{-k}
K^{\left(\frac{m}{2}-k\right)^2}\cdot
\binom mk_K \cdot\frac{\prod_{i=m+1-k}^m (QK^i-\epsilon)}{\prod_{i=1}^k (\epsilon K^i-Q)}}
\label{Mka=1}
\ee
\be\label{M01}
{\rm det} (M^{(0)})=(-)^{(m+4)(m^2+2m+3)\over 2}
K^{\frac{m(m+1)(2m-5)}{24}+\frac{1}{4}\left[\frac{m}{2}\right]}
(Q-\epsilon)^{\left[\frac{m}{2}\right]} \prod_{i=1}^m (K^i-1)^{m-i}
\prod_{i=1}^{m} (\epsilon K^i-Q)^{\left[\frac{m+2-i}{2}\right]}
\prod_{i=1}^{m} (Q K^i-\epsilon)^{\left[\frac{m-i}{2}\right]}
\ee
where $\binom mk_K$ are $K$-binomial coefficients\footnote{We define the $K$-number as $[n]_K=\displaystyle{{K^n-1\over K-1}}$ and the $q$-factorial as $[n]_K!=\prod_{i=1}^n[i]_K$.} and $[\ldots]$ denotes the integer part of the number.
The coefficients $\psi^{(1)}_{m,k}$ are given by the $k$-dependent expression in the box,
which is familiar from \cite{MMP2,MMP3}, while the (complicated) coefficient ${\rm det} (M^{(0)})$ is an inessential common normalization being independent on $k$.

\subsection{$a=2$}

Since we already know that no reasonable answer can be expected for generic $\epsilon$ for $a>1$,
from now on we substitute $\epsilon=e^{\frac{2\pi i}{a}}$. However, already in the case of $a=2$, the determinants are not that simple. In particular, they do not factorize at arbitrary $m$ and $k$. Indeed \cite{MMP1},
\be\label{Mka=2}
{\rm det} (M^{(k)}) &=& (-1)^k{\rm det} (M^{(0)})
\times
\ee
\be
\boxed{\times
\sum_{max(0,k-m)}^{[{k\over 2}]}(-1)^{k+r}K^{r(r-1)-2mr-(k-2r)(2m+1)}Q^{k-2r}{[m+k-2r]_{K^2}!\over [k-2r]_{K^2}![m-k+r]_{K^2}![r]_{K^2}!}
{\prod_{i=1}^r{Q^2K^{2(m-i+1)}-1\over K^2-1}\over\prod_{i=1}^{k-r}{Q^2K^{-2i}-1\over K^2-1}}
}\nn
\ee
\be\label{M02}
{\rm det} (M^{(0)}):&=&(-1)^{m(m-1)\over 2}\cdot 2^m\cdot K^{{1\over 4}m(2m^2+3m-1)+{1\over 2}[{m\over 2}]}\times\\
&\times&\prod_{s=1}^2(Q-\epsilon^s)^{[{m\over 2}]}\prod_{i=1}^m(\epsilon^sK^i-Q)^{[{m+2-i\over 2}]}
\prod_{i=1}^m(QK^i-\epsilon^s)^{[{m-i\over 2}]}\prod_{i=1}^{m-1}(K^i-\epsilon^s)^{m-i}\prod_{i=1}^{m-1}(\epsilon^sK^i-1)^{m-i}=\nn\\
&=&(-1)^{m^2(m-1)\over 2}\cdot 2^m\cdot K^{{1\over 4}m(2m^2+3m-1)+{1\over 2}[{m\over 2}]}\times\nn\\
&\times&(Q^2-1)^{[{m\over 2}]}\prod_{i=1}^m(K^{2i}-Q^2)^{[{m+2-i\over 2}]}
\prod_{i=1}^m(Q^2K^{2i}-1)^{[{m-i\over 2}]}\prod_{i=1}^{m-1}(K^{2i}-1)^{2(m-i)}\nn
\ee
where $\epsilon=-1$. Thus, there is an additional sum in this case.

\subsection{Generic $a$}

In the case of generic $a$, the expression fast becomes very complicated as $k$ grows. In particular, it is no longer factorizable (which we already see in the $a=2$ case) and, generally, contains several sums. However, there are some particular cases, when the determinant still factorizes. The simplest one is ${\rm det} (M^{(0)})$, which can be rewritten in the form
\be\label{M0}
{\rm det} (M^{(0)})=(-i)^{\Big({a^2\over 2}+{a\over 2}+1\Big)m^2}a^{\frac{am}{2}} K^{-\frac{am(m+1)(4am-a+3)}{12}}\prod_{s=1}^a  \prod_{n=0}^{m+a-1}
\left\{\prod_{i=2+(a-1)m+2n}^{(a+1)m-2n}(\epsilon^sK^i- u)\prod_{i=1 }^{m-1-n }(K^i-\epsilon^s)^a\right\}=\nn\\
=(-i)^{\Big({a^2\over 2}+{a\over 2}+1\Big)m^2}(-1)^{(a+1)m(m+1)\over 2}a^{\frac{am}{2}} K^{-\frac{am(m+1)(4am-a+3)}{12}}\prod_{n=0}^{m+a-1}
\left\{\prod_{i=2+(a-1)m+2n}^{(a+1)m-2n}(K^{ai}- u^a)\prod_{i=1 }^{m-1-n }(K^{ia}-1)^a\right\}
\ee
where $\epsilon=\exp\Big({2\pi i\over a}\Big)$, and we used that $\prod_{s=1}^a (X-\epsilon^s)=(X^a-1)$. Thus, one can see that only $K^a$ and $u^a$ (i.e. $q$ and $q^{am+\lambda}$) enters the result for the determinant. This phenomenon persists in all determinants ${\rm det} (M^{(k)})$ and, hence, in $\frac{\psi^{(a)}_{m,k}}{\psi^{(a)}_{m,0}}$. In the particular cases of $a=1,2$, expression (\ref{M0}) reduces to formulas (\ref{M01}), (\ref{M02}).

The next simplest case is at $k=am$. It is again factorized at any $a$ and $m$, contrary to what one could expect from \cite{MMP2}, the answer reads
\be\label{am}
{\rm det} (M^{(am)})={\rm det} (M^{(0)})\cdot
K^{\frac{am(m+1)}{2}}
\left( \frac{\prod_{i=(a-1)m}^{am-1} (K^{ai}-u^a)}{\prod_{i=am+1}^{(a+1)m} (K^{ai}-u^a)} \right)
\ee
This expression is certainly consistent with the boxes in (\ref{Mka=1}) and (\ref{Mka=2}) at $k=m$.

Also factorized are
\be
{\rm det} (M^{(1)})=-{\rm det} (M^{(0)})\cdot
\left(\frac{1-u }{ K^{m+1}-u}\right)^{\delta_{a,1}}\cdot N_{a,m}^+\\
N_{a,m}^+:=u^{a-1}K \frac{(K^{am}-1)(K^{a(m+1)}-1)}{(K^a-1)(K^{a(am+1)}-u^a)}
\label{psik=1}
\ee
and
\be\label{psidopk=1}
{\rm det} (M^{(am-1)})&=&(-1)^{am-1}{\rm det} (M^{(0)})\times (-1)^{m+1}\left(\frac{K^{2m}-u}{u(K^{m+1}-1)}\right)^{\delta_{a,1}}\cdot N_{a,m}^-\\
N_{a,m}^-:&=&u K^{(a-1)(am-1)+{am(m-1)\over 2}}\cdot\frac{(K^{am}-1)(K^{a(m+1)}-1)}{(K^a-1)(K^{a(am+1)}-u^a)}\cdot
\frac{\prod_{i=(a-1)m}^{am-2}(K^{ai}-u^a)}{\prod_{i=am+2}^{(a+1)m}(K^{ai}-u^a)}\nn
\ee
with $\epsilon=\exp\Big({2\pi i\over a}\Big)$.

Factorization breaks down for other values of $k$,
it grows with deviation of $k$ from the ends
at $k=0$ and $k=am$.
Another peculiarity is the division of the region $0\leq k \leq ma$ by points $k=ja$, $j=1,\ldots,m-1$ into segments,
with a peculiar behaviour in each of them.

For instance, for $0<k<a$ one has a combination of min($m,k$) factorized terms:
\be
{\rm det} (M^{(k)})&=&(-1)^{k}{\rm det} (M^{(0)})\times u^{a-k}K^{k^2+ma^2(k-1)} N_{a,m}^+
\cdot{{\cal P}_{k,m,a}^+(K)\over u^a-K^{a(am+2)}}
\ee
\be
{\cal P}_{1,m,a}^+&=&u^a-K^{a(am+2)}\nn\\
{\cal P}_{k,1,a}^+&=&u^a-K^{a(a+2)}\nn\\
{\cal P}_{k,2,a}^+&=&u^a(1+K^{2a}-K^{(k+1)a})-K^{2a^2+(k+1)a}\nn\\
\ldots\nn
\ee

In variables $q=K^a$ and $\Lambda:=q^\lambda=u^aK^{-a^2m}$, this formula is rewritten as

\bigskip

\hspace{-.6cm}\fbox{\parbox{17cm}{
\be\label{main1}
\underline{0<k<a}&:&\nn\\
{\rm det} (M^{(k)})&=&(-1)^{k}{\rm det} (M^{(0)})\times \Lambda^{1-\frac{k}{a}}q^{\frac{k^2}{a}-mk}
\frac{(q^m-1)(q^{m+1}-1)}{(q-1)\prod_{i=1}^k (q^i-\Lambda)}\times\nn\\
&\times&\sum_{j=0}^{\infty} \left\{q^{j(j+1)}\Lambda^j \prod_{i=1}^{k-j-1}(q^{m+i}-\Lambda) \prod_{i=1}^j \left((q^{k-i}-1)
\frac{q^{m-i}-1}{q^{i+1}-1}\frac{q^{m-i+1}-1}{q^{i}-1}\right)
\right\}
\ee
}}

\bigskip

In fact, this sum is finite: it is automatically cut-off at $j=$min$(m,k)-1$.
In the variables $q$ and $\Lambda$, the formulas are essentially independent of $a$: {\bf an $a$-dependence remains only in a simple overall factor}.

Similarly, in the last segment $(m-1)a< k < ma$, the answer is also a combination of min($m,k$) factorized items (note that we redefine $k\to ma-k$, i.e. $1\le k< a$):
\be
{\rm det} (M^{(am-k)})&=&(-1)^{am-k}{\rm det} (M^{(0)})\times (-1)^{m+1} u^{k-1}K^{(k-1)(k-a+1-ma)} N_{a,m}^-
\cdot{{\cal P}_{k,m,a}^-(K)\over u^a-K^{a(am-2)}}
\ee
where
\be
{\cal P}_{1,m,a}^-&=&u^a-K^{a(am-2)}\nn\\
{\cal P}_{k,1,a}^-&=&u^a-K^{a(a-2)}\nn\\
{\cal P}_{k,2,a}^-&=&u^a+K^{2a^2}\Big(1-K^{-(k-1)a}-K^{-(k+1)a}\Big)\nn\\
\ldots\nn
\ee
Using the variables $q$ and $\Lambda$, one can conveniently rewrite this formula in the form

\bigskip

\hspace{-.6cm}\fbox{\parbox{17cm}{
\be\label{main2}
\underline{0<k<a}&:&\nn\\
{\rm det} (M^{(am-k)})&=&(-1)^{am-k}{\rm det} (M^{(0)})\times \Lambda^{k\over a}
q^{{k^2\over a}+1-m-k} {(q^m-1)(q^{m+1}-1)\over (q-1)(\Lambda q-1)}\prod_{i=1}^m{q^i\Lambda-1\over q^i-\Lambda}\times\nn\\
&\times&\sum_{j=0}^{\infty}\left\{q^{j(j+2-k-m)}\prod_{i=1}^j\left({(q^{m+i+1}-1)(q^{m-i}-1)\over(q^{i+1}-1)(q^i-1)}
\cdot{(q^{k-i}-1)\over (q^{i+1}\Lambda-1)}\right)\right\}
\ee
}}

\bigskip

The sum is again finite: it is automatically cut-off at $j=$min$(m,k)-1$. And again, an $a$-dependence remains only in a simple overall factor. However, in order to test higher $k$, one has to choose higher $a$, since $k<a$.

\section{Phase transitions at $k=na$}

$M^{(k)}$ are defined as principal minors of the (rectangular) matrix obtained by eliminating
the $k$-th {\it column}.
Dependence on columns (on $k$ in (\ref{Eq})) is smooth, but that on {\it rows} (on $l$ in (\ref{Eq})) is not:
it changes at $l = integer\times a$.
Somehow this apparent discontinuity
in the horizontal direction is converted into discontinuity
of $M_k$ in the vertical direction, which we observe when looking at $M^{(k)}$.

In fact, the very existence of jumps is obvious from the very beginning: formula (\ref{main1}) can not be correct up to $k=am$, since, on one hand, it basically does not depend on $a$, and, on the other hand, it does not match the correct answer at $k=am$ (\ref{am}).

Indeed, for instance, at $k=a$, there is a jump: to the expression (\ref{main1}), an additional term (independent of $\Lambda$) is added:
\be
\Delta\Big[{\rm det} (M^{(a)})\Big]=(-1)^{a+1}{\rm det} (M^{(0)})\times q{(q^m-1)\over (q-1)}
\ee
i.e. the full answer is
\be
(-1)^{a} \frac{{\rm det} (M^{(a)})}{{\rm det} (M^{(0)})} =
\ee
\be
\!\!\!\!\!\!\!\!\!\!\!\!= \left[q^{a(1-m)}
\frac{(q^m-1)(q^{m+1}-1)}{(q-1)\prod_{i=1}^a (q^i-\Lambda)}
\sum_{j=0}^{\infty} \left\{q^{j(j+1)}\Lambda^j \prod_{i=1}^{a-j-1}(q^{m+i}-\Lambda) \prod_{i=1}^j \left((q^{a-i}-1)
\frac{q^{m-i}-1}{q^{i+1}-1}\frac{q^{m-i+1}-1}{q^{i}-1}\right)\right\}\underline{-q{(q^m-1)\over (q-1)}}
\right]=\nn
\ee

\be
= {(q^m-1)\over (q-1)}\left[\underline{-q}+\frac{q^{a(1-m)}(q^{m+1}-1)}{\prod_{i=1}^a (q^i-\Lambda)}
\sum_{j=0}^{\infty} \left\{q^{j(j+1)}\Lambda^j \prod_{i=1}^{a-j-1}(q^{m+i}-\Lambda) \prod_{i=1}^j \left((q^{a-i}-1)
\frac{q^{m-i}-1}{q^{i+1}-1}\frac{q^{m-i+1}-1}{q^{i}-1}\right)\right\}
\right]\nn
\ee
The jump is underlined.

\bigskip

Similarly, the jump at the boundary of the last segment, at $k=a(m-1)$ is
\be
\Delta\Big[{\rm det} (M^{(a(m-1))})\Big]={\rm det} (M^{(0)})\times(-1)^{a(m-1)+1}q{(q^m-1)\over(q-1)}\prod_{j=1}^m{(q^i\Lambda-1)\over(q^i-\Lambda)}
\ee
which adds to (\ref{main2}) so that the full answer is
\be
&(-1)^{a(m-1)}&{{\rm det} (M^{(a(m-1))})\over{\rm det} (M^{(0)})}= \left[\Lambda
q^{1-m} {(q^m-1)(q^{m+1}-1)\over (q-1)(\Lambda q-1)}\prod_{i=1}^m{q^i\Lambda-1\over q^i-\Lambda}\times\right.\nn\\
&\times&\left.\sum_{j=0}^{\infty}\left\{q^{j(j+2-a-m)}\prod_{i=1}^j\left({(q^{m+i+1}-1)(q^{m-i}-1)\over(q^{i+1}-1)(q^i-1)}
\cdot{(q^{a-i}-1)\over (q^{i+1}\Lambda-1)}\right)\right\}\underline{-q{(q^m-1)\over(q-1)}\prod_{i=1}^m{(q^i\Lambda-1)\over(q^i-\Lambda)}}\right]=\nn\\
&=&(-1)^{a(m-1)}{\rm det} (M^{(0)})\times {(q^m-1)\over (q-1)}\prod_{i=1}^m{q^i\Lambda-1\over q^i-\Lambda}\Big[\underline{-1}+\nn\\
&+&{\Lambda
q^{1-m}(q^{m+1}-1)\over (\Lambda q-1)}\left.\sum_{j=0}^{\infty}\left\{q^{j(j+2-a-m)}\prod_{i=1}^j\left({(q^{m+i+1}-1)(q^{m-i}-1)\over(q^{i+1}-1)(q^i-1)}
\cdot{(q^{a-i}-1)\over (q^{i+1}\Lambda-1)}\right)\right\}\right]
\ee

Going further to $a<k<2a$, another jump term adds, and similarly for the segment $ma-2a<k<ma-a$, etc.

One can continue constructing the answers for other segments, however, they become quite involved. Hence, we stop their description at this point, and switch to a short{ discussion.

\section{Discussion: why so involved?
}

One can wonder how it happens that an immediate and simple exercise in linear algebra is so complicated
that may require a scientific paper to discuss it.
Let us localize the problem.

In abstract notation, one needs to solve the system
\be
A_{\alpha\beta} \psi_\beta = A_{\alpha 0}
\ee
i.e. to find $\psi = A^{-1}A_0$ for the matrix of the form
\be
A_{\alpha\beta} =  {\bf 1} - x^\alpha y^{\alpha\beta}
\ee
where ${\bf 1}$ denotes the matrix filled by unities.

In practice, $\alpha$ is a multi-index, $x^\alpha = u^{-i} \epsilon^{-s}$, which makes an additional complication,
but it is not the main one.
Much more important is that  ${\bf 1}$ is not the unit matrix $\delta_{\alpha\beta}$,
and it has rank $1$.
Therefore it is impossible, say, to expand the inverse matrix $A^{-1}$ in powers of $y$.
The variables $x$ do not matter for this, problematic is already the inverse matrix of
a symmetric matrix $ {\bf 1} - y^{\alpha\beta}$.
Already in the $4\times 4$ case, it is
\be
\frac{1}{(y-1)(y^2-1)(y^3-1)(y^4-1)}\times\nn\\
\ee
\be
\times\left(\begin{array}{cccc}
-{y^2(y^4+y^2+y+1)\over (y+1)(y^2+1)} & {y^2-y+1\over y(y+1)^2(y^2+y+1)(y^2+1)} & -{y^4+y^3+y^2+1\over y^3(y+1)(y^2+1)}&{1\over y^3(y+1)(y^2+1)} \\
\\
{y^2-y+1\over y(y+1)^2(y^2+y+1)(y^2+1)} & -{y^6+y^4+2y^3+y^2+1\over y^4(y^2+1)(y^2+y+1)} & {y^2-y+1\over y^5(y+1)^2(y^2+y+1)(y^2+1)} &
 -{1\over y^5(y^2+1)(y^2+y+1)} \\
\\
-{y^4+y^3+y^2+1\over y^3(y+1)(y^2+1)} & {y^2-y+1\over y^5(y+1)^2(y^2+y+1)(y^2+1)} & -{y^4+y^2+y+1\over y^6(y+1)(y^2+1)}&{1\over y^6(y+1)(y^2+1)} \\
\\
{1\over y^3(y+1)(y^2+1)}& -{1\over y^5(y^2+1)(y^2+y+1)}&{1\over y^3(y+1)(y^2+1)} &-{1\over y^6}
\end{array}\right)
\ee
and even this oversimplified example illustrates appearance of somewhat ugly structures.
Actually they become a little better after convolution with the vector $A_{\alpha,0}$,
but remain somewhat non-trivial, which is reflected in explicit formulas in
the previous sections.

At the same time, solution of this particular problem of linear algebra is of importance,
because it is {\bf an underlying structure in the theory of Macdonald polynomials} and its generalizations.
It remains tricky basically because of the lack of an explicit inversion
of the simply-looking matrix ${\bf 1}-y^{\alpha\beta}$.

\section{Conclusion}

The goal of this short note is to explain a non-triviality of the seemingly simple
Chalykh equations, complementing the {\it triad} \cite{MMP3}, which
involves more sophisticated objects of Macdonald theory
including the Noumi-Shiraishi non-polynomial extension \cite{NS} related to the concept of {\it mother function} \cite{Smirnov,AKMM1} (see also earlier papers \cite{Ok}).
Despite being just linear equations, solvable with the elementary-school Cramer's rule,
the Chalykh equations turn expressed through unexpectedly difficult determinants.

Though an elliptic counterpart of the Noumi-Shiraishi functions is known \cite{FOS,AKMM2,AKMM3},
an elliptic deformation of the Chalykh equations, which could seem simple \cite{MMPZ1} appears to be ambiguous and also requires additional insights.
Note that the lift of the Noumi-Shiraishi functions to the Shiraishi functions \cite{Shi,LNS} introducing another ellipticity also satisfies a set of Chalykh equations though this time they are not enough to fix a unique (up to normalization) solution \cite{MMPZ2}.

\section*{Acknowledgements}

This work is supported by the RSF grant 24-12-00178.

\end{document}